\documentclass[graybox]{svmult}

\usepackage{type1cm}        
\usepackage{makeidx}         
\usepackage{graphicx}        
\usepackage{multicol}        
\usepackage[bottom]{footmisc}

\usepackage{newtxtext}       %
\usepackage[varvw]{newtxmath}       

\makeindex             

\begin{document}

\title*{Over-Measurement Paradox: Suspension of Thermonuclear
    Research Center and Need to Update Standards}
\titlerunning{Over-Measurement Paradox:
Thermonuclear
    Research Center and Standards}
\author{Hector Reyes, Saeid Tizpaz-Niari, and Vladik Kreinovich}
\institute{
Hector Reyes, Saeid Tizpaz-Niari, and Vladik Kreinovich \at
Department of Computer Science, University of Texas at El Paso\\500
W. University, El Paso, Texas 79968,
USA\\\email{hareyes2@miners.utep.edu,
saied@utep.edu,vladik@utep.edu}}
\maketitle

\abstract*{In general, the more measurements we perform, the more
information we gain about the system and thus, the more adequate
decisions we will be able to make. However, in situations when we
perform measurements to check for safety, the situation is
sometimes opposite: the more additional measurements we perform
beyond what is required, the worse the decisions will be: namely,
the higher the chance that a perfectly safe system will be
erroneously classified as unsafe and therefore, unnecessary
additional features will be added to the system design. This is not
just a theoretical possibility: exactly this phenomenon is one of
the reasons why the construction of a world-wide thermonuclear
research center has been suspended. In this paper, we show that the
reason for this paradox is in the way the safety standards are
formulated now -- what was a right formulation when sensors were
much more expensive is no longer adequate now when sensors and
measurements are much cheaper. We also propose how to modify the
safety standards so as to avoid this paradox and make sure that
additional measurements always lead to better solutions.}

\abstract{In general, the more measurements we perform, the more
information we gain about the system and thus, the more adequate
decisions we will be able to make. However, in situations when we
perform measurements to check for safety, the situation is
sometimes opposite: the more additional measurements we perform
beyond what is required, the worse the decisions will be: namely,
the higher the chance that a perfectly safe system will be
erroneously classified as unsafe and therefore, unnecessary
additional features will be added to the system design. This is not
just a theoretical possibility: exactly this phenomenon is one of
the reasons why the construction of a world-wide thermonuclear
research center has been suspended. In this paper, we show that the
reason for this paradox is in the way the safety standards are
formulated now -- what was a right formulation when sensors were
much more expensive is no longer adequate now when sensors and
measurements are much cheaper. We also propose how to modify the
safety standards so as to avoid this paradox and make sure that
additional measurements always lead to better solutions.}

\section{What Is Over-Measurement Paradox}

\noindent{\bf General case: the more measurements, the better.} Most of our knowledge about the world comes from measurements; see, e.g., \cite{Rabinovich 2005}. Each measurement provides is with an additional information about the world -- and once we have a sufficient number of measurements of the same system, we may be able to find the equations that describe the dynamics of this system and thus, to get even more additional information that was hidden in the original measurements.

The more measurements we perform, the more information we gain about the system, the more accurate our estimates, and thus, the better will be our decisions. From this viewpoint:
\begin{itemize}
\item the more measurements we perform,
\item the better.
\end{itemize}
We only expect one limitation on the number of measurements -- the financial one. Indeed, at some point, after we have performed a large number of measurements, we get a very accurate picture of the measured system. Decisions based on this picture are close to optimal, and a very small expected increase in optimality may not be worth spending money on additional measurements.
\medskip

\noindent{\bf Over-measurement paradox: case study.} Most of our energy comes from the Sun. In the Sun, as in most stars, energy is generated by the thermonuclear synthesis, when protons -- i.e., nuclei of Hydrogen (H) -- combine together to form nuclei of Helium (He). This is a very efficient way of generating energy, a way that does not lead to pollution or other side effects. The majority of physicists believe that this is a way to get energy for our civilization: instead of relying on direct or indirect energy from the thermonuclear reaction inside the Sun, why not use the same reactions ourselves -- this will be a very effective and clean idea.

The idea is theoretically feasible, but technically, this is a very difficult task. Researchers and engineers all over the world have been working on it since the 1950s. To speed up the process, researchers from 35 major world countries decided to join efforts, and allocated \$65 billion dollars to build an international research center where specialists from all the world will work on this topic. This project is named ITER -- this is both:
\begin{itemize}
\item an abbreviation of International Thermonuclear Experimental Reactor and
\item the Latin word meaning ``the way"; see, e.g., \cite{ITER}.
\end{itemize}

The problem is that as of now, this project is suspended, and one of the main reasons for this suspension is over-measurement; see, e.g., \cite{Kramer 2022}. In a nutshell, the requirement was that, to guarantee safety, the level of danger -- e.g., the level of radiation -- was supposed to be below the safety threshold at a certain number of locations and scenarios.
\begin{itemize}
\item The current design does satisfy this criterion.
\item However, the designers decided to be thorough and simulated more measurement situations.
\end{itemize}
Unfortunately, some of the expected measurement results exceed the threshold. As a result, the whole project is in suspension. Making sure that all future measurements satisfy the criterion would require a drastic redesign and a drastic further increase in the cost of the whole project -- so drastic that it is doubtful that this additional funding will appear, especially in the current economic situation.

Why is it a paradox? If the designers did not perform these additional measurements, the design would have been approved and the project would have started. So in this case, additional measurements made the situation much worse -- not only for the researchers, but for the humankind as a whole. This is a clear situation where additional measurements do not help at all.
\medskip

\noindent{\bf But is it really a paradox?}
Maybe it is good that the project stopped -- maybe additional measurements revealed that the original design was unsafe?
\medskip

\noindent{\bf What we do in this paper.} In this paper, we analyze the situation from the general measurement viewpoint and come up with several conclusions.
\begin{itemize}
\item first, we show that this situation is, in principle, ubiquitous: a similar problem will surface in many other projects, including those that have already been approved and designed and seem to function OK;
\item second, although it may look that the problem is caused by insufficient safety of the original design, we show that this is not the case: practically any design, no matter how safe, will fail the currently used criteria if we perform sufficiently many measurements;
\item finally, we propose a natural suggestion on now to change the corresponding standards so as to avoid such unfortunate situations.
\end{itemize}

\section{Analysis of the Problem}

\noindent{\bf Let us formulate the situation in precise terms.} We are interested in studying states of different systems. A usual way to describe each state is by describing the values of the corresponding quantities at different locations and at different moments of time.

Usually, specifications include constraints on the values of some of these quantities. These may be constrains on the radioactivity level, constraints on concentration of potentially harmless chemicals, on the temperature, etc.

In all these cases, a typical constraint is that the value of some quantity $q$ should not exceed some threshold $q_0$: $q\le q_0$.
\medskip

\noindent{\bf How can we check this constraint: seemingly natural idea.} In the ideal world, we should be able to measure the value $q(x,t)$ at all possible spatial locations $x$ and for all possible moments of time $t$, and check that all these values do not exceed $q_0$.

Of course, in real life, we can only perform finitely many measurements. So, a seemingly natural idea is to perform several measurements, and to check that all measurement results $q_1,\ldots,q_n$ do not exceed $q_0$. However, it is known that this seemingly natural idea can lead to dangerous consequences; see, e.g., \cite{Rabinovich 2005}. Let us explain why.
\medskip

\noindent{\bf Why the above seemingly natural idea is dangerous.} The actual value of the quantity $q$ depends on many factors which are beyond our control. For example, the actual radioactivity level at a given location is affected by the natural radioactivity level at this location -- the level that can change based, e.g., on weather conditions, when wind brings matter from neighboring areas where this natural level is somewhat higher. There are many small independent factors affecting the actual value of the quantity $q$.

In addition, the measurement result is somewhat different from the actual value of the measured quantity; see, e.g., \cite{Rabinovich 2005}. We may be able to get rid of major sources of such measurement errors, but there are always a lot of small independent factors that lead to small changes of the measurement results.

Because of both types of random factors, the measured value differs from its locally-average level, and this difference is the result of a joint effort of a large number of small independent factors. It is known (see, e.g., \cite{Sheskin 2011}) that such a joint effect is usually well described by a normal (Gaussian) distribution. To be more precise:
\begin{itemize}
\item What is known is that in the limit, when the number $N$ of small independent random factors increases (and the size of each factor appropriately decreases), the probability distribution of the joint effect of all these factors tends to the normal distribution -- which thus appears as the limit of the actual distributions when $N$ increases.
\item By definition of the limit, this means exactly that when the number $N$ of factors is large -- and in many practical situations it is large -- the actual distribution is very close to normal.
\end{itemize}
So, with high accuracy, we can safely assume that this distribution is normal.

This assumption explains why the above seemingly natural idea is dangerous. Indeed, what we have is several measurement results $q_1,\ldots,q_n$, i.e., in effect, several samples from the normal distribution. Usually, measurement errors corresponding to different measurements are practically independent -- and the same can be said about the random factors affecting the value of the quantity $q$ at different spatial locations and at different moments of time. From this viewpoint, what we observe are $n$ independent samples from a normal distribution.

If we only require that $q_i\le q_0$, we thus require that $\max(q_1,\ldots,q_n)\le q_0$. Usually, our resources are limited, so we try to make the minimal effort to satisfy the requirements. In other words, when we institute more and more efficient filters -- thus slowly decreasing the value $q_i$ -- and finally, reach the condition $\max(q_1,\ldots,q_n)\le q_0$, we stop and declare this design to be safe.
\begin{itemize}
\item
We start with the design for which $\max(q_1,\ldots,q_n)>q_0$.
\item So the first time when we satisfy the desired constraint $\max(q_1,\ldots,q_n)\le q_0$ is when we get $$\max(q_1,\ldots,q_n)=q_0.$$
\end{itemize}
This again may sound reasonable, but it is known that the probability that the next random variable will exceed the maximum $\max(q_1,\ldots,q_n)$ is proportional to $1/(n+1)$. So:
\begin{itemize}
\item even if we perform 40 measurements -- and this is, e.g., what measurement theory requires for a thorough analysis of a measuring instrument (see, e.g., \cite{Rabinovich 2005}),
    \item we get a $1/40\approx 2.5\%$ probability that next time, we will go beyond the safety threshold.
\end{itemize}
This is clearly {\it not} an acceptable level of safety -- especially when we talk about serious, potentially deadly dangers like radioactivity or dangerous chemicals.
\medskip

\noindent{\bf So what can be done to avoid this danger.} To simplify our analysis, let us suppose that the mean value of $q$ is 0. This can always be achieved if we simply subtract the actual mean value from all the measurements result, i.e., for example, consider not the actual radioactivity level, but the excess radioactivity over the average value of the natural radioactivity background.

In this case, measurement results $q_1,\ldots,q_n$ form a sample from a normal distribution with 0 mean and some standard deviation $\sigma$.
\begin{itemize}
\item Of course, no matter how small $\sigma$, the normally distributed random variable always has a non-zero probability of being as large as possible -- since the probability density function of a normal distribution is always positive, and never reaches 0.
\item So, we cannot absolutely guarantee that all future values of $q$ will be smaller than or equal to $q_0$.
\item We can only guarantee that the probability of this happening is smaller than some given probability $p_0$, i.e., that $${\rm Prob}(q>q_0)\le p_0.$$
\end{itemize}
So, to drastically decrease the probability of a possible disaster -- from the unsafe $2.5\%$ to the much smaller safety level $p_0\ll 2.5\%$:
\begin{itemize}
\item instead of the original threshold $q_0$,
\item we select a smaller threshold $\widetilde q_0<q_0$ that guarantees that the conditional probability of exceeding $q_0$ is small:
$${\rm Prob}(q>q_0\,|\,\max(q_1,\ldots,q_n)\le \widetilde q_0)\le p_0.$$
\end{itemize}
In this case:
\begin{itemize}
\item if we have $n$ measurement $q_1,\ldots,q_n$ all below $\widetilde q_0$,
\item then we guarantee, with almost-1 probability $1-p_0$, that the next value will be below the actual danger threshold~$q_0$.
\end{itemize}
This value $\widetilde q_0$ depends on $q_0$ and on the number of measurements $n$:
\begin{itemize}
\item the larger $n$,
\item the larger the value $\widetilde q_0$.
\end{itemize}
When $n$ increases, this value tends to $q_0$.
\medskip

\noindent{\bf So what is included in the safety standard.} When safety standards are designed, one of the objectives is to make them easy to follow:
\begin{itemize}
\item We do not want practitioners -- who need to follow these standards -- to perform complex computations of conditional probabilities.
\item We need to give them clear simple recommendations.
\end{itemize}
From this viewpoint, the easiest to check if whether the measurement result satisfies a given inequality.

So, a reasonable way to set up the corresponding standard is to set up:
\begin{itemize}
\item the new threshold $\widetilde q_0$ and
\item the minimal necessary number of measurements $n$.
\end{itemize}
The standard then says that:
\begin{itemize}
\item if we perform $n$ measurements, and the results $q_1,\ldots,q_n$ of all these $n$ measurements do not exceed this threshold $\widetilde q_0$, then the situation is safe;
    \item otherwise, the situation is not safe, and additional measures need to be undertaken to make this situation safer.
\end{itemize}
\medskip

\noindent{\bf Resulting common misunderstanding.} The fact that safety standards provide such a simplified description -- and rarely mention actual threshold $q_0>\widetilde q_0$ -- makes most people assume that the critical value $\widetilde q_0$ provided by a standard is the actual danger level, so any situation in which a measured value exceeds $\widetilde q_0$ is unacceptable. This is exactly what happened in the above case study.

And this is wrong conclusion:
\begin{itemize}
\item if we perform a sufficiently large number of measurements,
\item we will eventually get beyond any threshold.
\end{itemize}
Indeed, according to the extreme value theory (see, e.g., \cite{Beirlant
2004,de Haan 2006,Embrechts 2012,Gumbel 2004,Lorkowski 2015}), for normal distribution with mean 0 and standard deviation $\sigma$, the average value $A_n$ of the maximum $\max(q_1,\ldots,q_n)$ grows with $n$ as
$$A_n\sim \gamma\cdot\sqrt{2\ln(n)}\cdot \sigma,$$ where $\gamma\approx 0.5772$
is the Euler's constant
$$\gamma\stackrel{\rm def}{=}\lim_{n\to\infty} \left(\sum_{k=1}^n \frac{1}{k}-\ln(n)\right).$$
So, this mean value indeed grows with $n$.
\medskip

\noindent{\bf Why this problem surfaces only now?} Gaussian distributions was invented by Gauss in early 19 century, measurements have been performed since antiquity, so why is this problem surfacing only now? Why did not it surface much earlier?

The main reason, in our opinion, is that, until recently:
\begin{itemize}
\item sensors were reasonably expensive -- especially accurate ones -- and the cost of measurements was non-negligible;
\item in this case, while in principle, it was possible to perform more measurement than requires for safety testing, this would have led to useless costs.
\end{itemize}
Lately, however:
\begin{itemize}
\item sensors have become very cheap: kids buy them to make robots, the cheapest cell phones have very accurate sensors of positions, acceleration, etc.;
    \item as a result, it is reasonably inexpensive to perform many more measurements than required;
    \item and, as we have mentioned, as a result, in situations that would previously -- based on only the required number of measurements -- would be classified as safe, now we get values exceeding the threshold $\widetilde q_0$ provided by the standard -- and thus, we end up classifying perfectly safe situations as unsafe.
\end{itemize}

\section{So What Do We Propose}

\noindent{\bf What is the problem now: summarizing our findings.} The reason why we have the over-measurement paradox is that current safety standards usually list only two numbers:
\begin{itemize}
\item the recommended threshold $\widetilde q_0$ and
\item the recommended number of measurements $n$.
\end{itemize}
The idea that the results of all the measurements must not exceed $\widetilde q_0$ for the situation to be considered safe.

The problem is that the recommended threshold $\widetilde q_0$ is actually {\it not} the safety threshold $q_0$, it is smaller than the safety threshold -- smaller so that for the prescribed number of measurements $n$, we would guarantee that:
\begin{itemize}
\item for all future values,
\item the probability to exceed the real safety threshold $q_0$ should be smaller than the desired small value $p_0$.
\end{itemize}

When, in an actually safe situation, in which the probability to exceed $q_0$ does not exceed $p_0$, we perform more measurements than recommended, then it is eventually inevitable that some of them will be larger than the recommended threshold $\widetilde q_0$ -- even though they will still, with almost-1 probability, be not larger than the actual danger threshold $q_0$. This leads to the following natural solution to the over-measurement problem.
\medskip

\noindent{\bf Proposed solution: we need to change the standards.} In addition to providing the two numbers $\widetilde q_0$ and $n$, we should provide the formula describing the dependence of the testing safety threshold $t(n')$ for different numbers $n'\ge n$ of actual measurements, so that for all $n'$, we should have
$${\rm Prob}(q>q_0\,|\,\max(q_1,\ldots,q_{n'})\le t(n'))\le p_0.$$
At least we should provide the value $t(n')$ for several different values $n'$, thus taking care of the cases when, due to thoroughness, practitioners will provide more measurements.

\section*{Acknowledgments}

This work was supported in part by the National Science Foundation
grants 1623190 (A Model of Change for Preparing a New Generation
for Professional Practice in Computer Science), and HRD-1834620 and
HRD-2034030 (CAHSI Includes), and by the AT\&T Fellowship in
Information Technology.

It was also supported by the program of the development of the
Scientific-Educational Mathematical Center of Volga Federal
District No. 075-02-2020-1478, and by a grant from the Hungarian
National Research, Development and Innovation Office (NRDI).


\begin{thebibliography}{9}

\bibitem{Beirlant 2004} J. Beirlant, Y. Goegevuer, J. Teugels, and
J. Segers, {\it Statistics of Extremes: Theory and Applications}, Wiley, Chichester, 2004.

\bibitem{de Haan 2006}
L. de Haan and A. Ferreira, {\it Extreme Value Theory: An Introduction}, Springer Verlag, Berlin, Hiedelberg, New York, 2006.

\bibitem{Embrechts 2012}
P. Embrechts, C. Kl\"uppelberg, and T. Mikosch,
{\it Modelling Extremal Events for Insurance and Finance}, Springer Verlag, Berlin, Heidelberg, New York, 2012.


\bibitem{Gumbel 2004} E. J. Gumbel, {\it Statistics of Extremes}, Dover Publ., New York, 2004.

\bibitem{ITER} ITER Project, https://www.iter.org/

\bibitem{Kramer 2022} D. Kramer, ``Further delays at ITER are certain, but
their duration isn't clear'', {\it Physics Today}, May 2022, pp. 20--22.

\bibitem{Lorkowski 2015} J. Lorkowski, O. Kosheleva, V. Kreinovich, and S.
  Soloviev, ``How design quality improves with increasing
  computational abilities: general formulas and case study of
  aircraft fuel efficiency", {\it Journal of Advanced Computational
  Intelligence and Intelligent Informatics (JACIII)}, 2015, Vol. 19,
  No. 5, pp. 581--584.

\bibitem{Rabinovich 2005} S. G. Rabinovich, {\it Measurement Errors and
Uncertainty: Theory and Practice}, Springer Verlag, New York, 2005.

\bibitem{Sheskin 2011}
D. J. Sheskin, {\it Handbook of Parametric and Nonparametric Statistical
Procedures}, Chapman and Hall/CRC, Boca Raton, Florida, 2011.

\end{thebibliography}
\end{document}